# ПЛАВЛЕНИЕ И КРИСТАЛЛИЗАЦИЯ В СЛОИСТОЙ ПЛЕНОЧНОЙ СИСТЕМЕ Ge-Bi

С.И. Богатыренко*, Н.Т. Гладких, С.В. Дукаров, А.П. Крышталь

*Научный физико-технологический центр МОН и НАН Украины, (Харьков)
Харьковский национальный университет им. В.Н. Каразина
Украина

Приведены результаты исследований плавления и кристаллизации в слоистой пленочной системе висмут – германий, создаваемой в вакууме путем последовательной конденсации компонентов при испарении их из независимых источников. Показано, что температура плавления в указанной системе понижается с уменьшением толщины пленки легкоплавкого компонента. Дифференциальный метод, использованный для регистрации температуры плавления, позволил найти значение эвтектической температуры $T_e$ = 542 К в системе Bi-Ge. Определены величина переохлаждения при кристаллизации ($\Delta T$ = 93 К) и краевой угол смачивания ($\theta$ = 68°) для островковых пленок висмута на аморфной германиевой подложке.

## ВВЕДЕНИЕ

Качественно новые задачи, появившиеся в последнее время в нанофизике и в современном материаловедении, значительно стимулировали интерес к процессам плавления и кристаллизации микрообъектов, внедренных в твердотельную матрицу [1 – 7]. Особый интерес в плане практических применений вызывают бинарные системы компоненты, которых практически полностью нерастворимы в твердом состоянии. Однако имеющиеся в литературе данные по изменению температур фазовых переходов вследствие влияния матрицы неоднозначны, поэтому в работе [8] был предложен новый метод исследования систем "частица-матрица", основанный на использовании конденсированных в вакууме пленок, а также разработан дифференциальный метод для регистрации незначительных изменений температур фазовых переходов жидкость – кристалл. Исследования ряда слоистых систем типа Al/(In, Sn, Bi, Pb)/Al, выполненные в работе [8] показали, что температура плавления пленки легкоплавкого элемента, находящегося в кристаллической матрице, т.е. между толстыми слоями Al, понижается с уменьшением ее толщины.

Данная работа продолжает цикл работ посвященных изучению фазовых переходов для систем, когда один из компонентов внедрен в твердотельную матрицу с более высокой температурой плавления. В настоящей работе выполнены исследования температур плавления и кристаллизации для висмута, находящегося в контакте с аморфной матрицей (конденсированные пленки Ge). Система Bi-Ge характеризуется фазовой диаграммой эвтектического типа с незначительной растворимостью в твердом состоянии и неограниченной – в жидком [9]. Для изучения особенностей фазовых переходов в исследуемой системе использовались разработанные ранее методики регистрации температур плавления [8] и крис-таллизации [10, 11] в конденсированных пленках.

## МЕТОДИКА ИССЛЕДОВАНИЙ

Методика препарирования образцов подробно описана в работе [8] и вкратце состояла в следующем. На протяженную полированную подложку из нержавеющей стали в вакууме ~$1\cdot10^{-6}$ мм рт.ст. конденсировалась пленка углерода толщиной примерно 20 нм для предотвращения взаимодействия материала подложки с исследуемой слоистой пленочной системой. Затем без нарушения вакуума на половину подложки (по ее ширине) с углеродной пленкой конденсировалась толстая (100 нм) пленка германия и сразу же вслед за нею на всю ширину подложки наносилась толстая пленка висмута. В ряде экспериментов на подложке препарировались трехслойные системы, в которых пленка висмута разной толщины находилась между толстыми пленками германия. При этом для контроля температуры всегда на подложке по ее ширине рядом с исследуемой слоистой системой конденсировалась толстая пленка висмута, температура плавления которой хорошо известна и которая служила реперной точкой, позволяющей определять даже незначительные изменения исследуемых температур в слоистой системе.

Для обеспечения необходимой последовательности нанесения слоев между подложкой и испарителями размещалась система подвижных разделительных экранов. После прекращения конденсации вдоль подложки создавался необходимый перепад температур. На всех образцах в направлении возрастания температуры визуально четко обнаруживается граница, которая соответствует температуре плавления легкоплавкого компонента (висмута). Это возможно благодаря тому, что висмут в жидком состоянии не смачивает углеродную пленку (угол смачивания и θ = 130°) и после расплавления пленка висмута собирается в отдельные сферические капли. Рядом на той же подложке, на другой половине по его ширине, видна граница, которая соответствует температуре контактного плавления в слоистой пленочной системе (рис. 1).

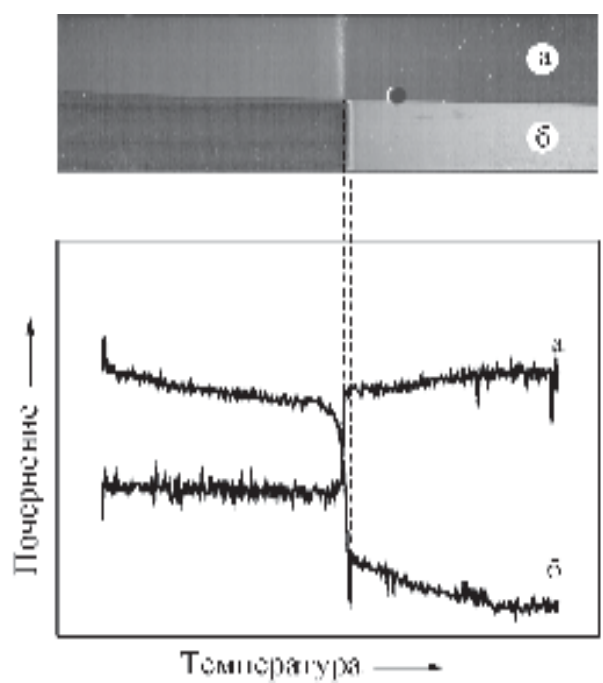

Рис. 1. Фотография подложки со слоистой пленочной системой Bi/Ge (а) и пленкой Bi (б) на аморфном углеродном подслое и соответствующие им фотометрические кривые. Толщина пленок Ge и Bi – 100 нм

Зная распределение температуры вдоль подложки и наблюдая за взаимным расположением границ, которые соответствуют температуре плавления чистого висмута (эта температура служит для контроля абсолютных значений температуры) и температуре контактного плавления в исследуемой пленочной системе, т.е. эвтектической температуре, можно с достаточной степенью надежности судить о величине температуры плавления эвтектики.

Определение температуры переохлаждения при кристаллизации Bi на аморфной германиевой пленке ($T_g$) выполнялось следующим образом. На протяженную подложку с предварительно нанесенным экранирующим слоем углерода при комнатной температуре конденсировалась пленка аморфного германия. Далее вдоль подложки создавался градиент температур в интервале 400 – 600 К и затем конденсировался висмут. Положение температуры $T_g$ определялось по изменению микроструктуры пленок выше и ниже ее [10, 11] и далее уточнялось с помощью электронно-микроскопических исследований.

## ЭКСПЕРИМЕНТАЛЬНЫЕ РЕЗУЛЬТАТЫ И ИХ ОБСУЖДЕНИЕ

Поскольку в литературе нет однозначных значений температуры плавления эвтектики, образованной висмутом с германием, а для конденсированных пленок, когда пленки германия являются аморфными, подобные данные отсутствуют, то сначала была проведена серия экспериментов для ее определения с помощью дифференциального метода. Для этого на половину подложки с углеродной пленкой препарировалась по всей ее длине двухслойная пленка указанных компонентов, а рядом, на другой половине, – толстая пленка висмута. Положение границ, соответствующих плавлению пленок, определялось путем фотометрирования снимков подложки. На приведенной фотографии подложки (рис. 1) четко видно, что температура плавления эвтектики ниже температуры плавления Bi и составляет 542 К.

Были проведены рентгеноструктурные исследования слоистой системы после плавления и охлаждения до комнатной температуры на рентгеновском дифрактометре Дрон-3М с использованием $K_\alpha$ излучения меди. Полученные значения параметров решетки висмута соответствуют табличным данным для чистого висмута, что свидетельствует об отсутствии заметной растворимости компонент в системе Bi-Ge.

Таким образом, из выполненных экспериментов следует, что температура плавления в конденсированных пленочных системах Bi/Ge, Ge/Bi/Ge, соответствующая эвтектической температуре системы Bi – аморфный Ge, на 2,5 К ниже известного из литературы значения для системы Bi – кристаллический Ge [9].

Для определения размерной зависимости эвтектической температуры в исследуемой системе была проведена серия экспериментов, в которых на одну подложку без нарушения вакуума наносились слоистые системы с разной толщиной пленки висмута. Использование специальной системы подвижных экранов позволяло получать на одной подложке, по ее ширине, в одном эксперименте до 5 слоистых пленочных систем с разным чередованием слоев и их толщин. Фотография подложки с такими слоистыми системам приведена на рис. 2. Видно что, в таких экс-

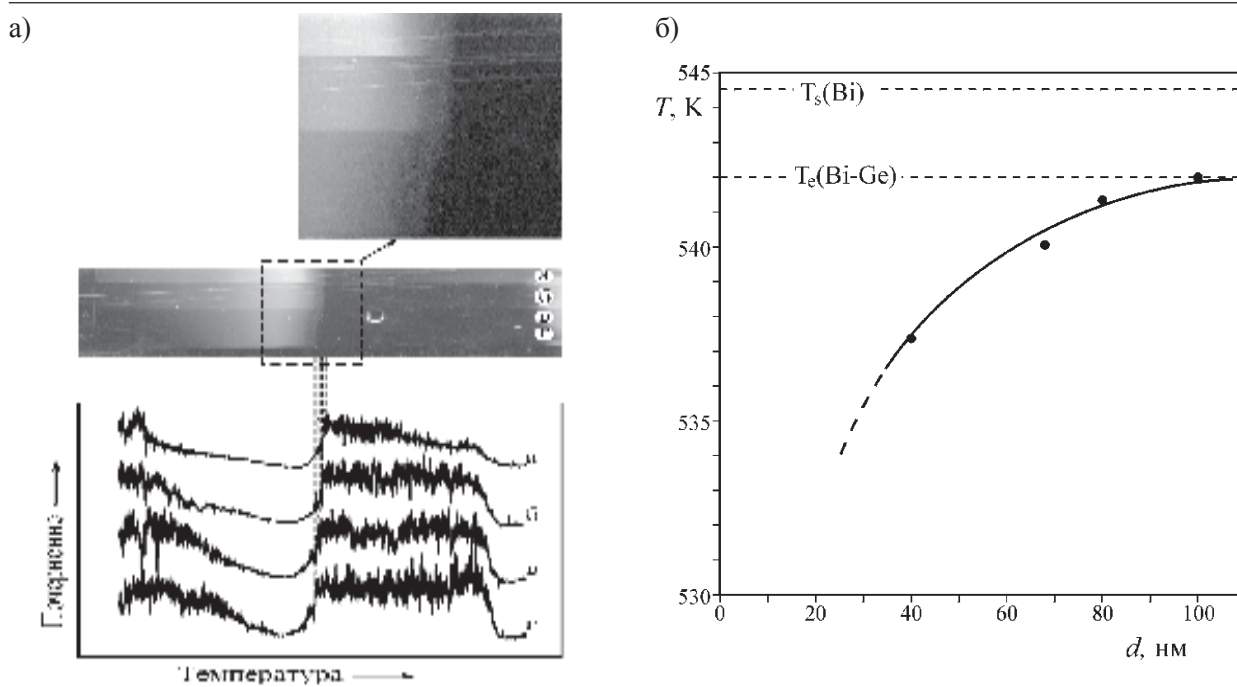

Рис. 2. Фотография подложки со слоистыми пленочными системами Ge/Bi/Ge при разных толщинах пленки висмута (а – 100 нм, б – 80 нм, в – 68 нм и г – 40 нм) (а) и соответствующая размерная зависимость температуры плавления (б) периментах на подложке визуально наблюдается сдвиг границы плавления в сторону меньших температур при уменьшении толщины пленок висмута, то есть температура плавления в слоистых системах Ge/Bi/Ge понижается с уменьшением толщины пленки легкоплавкого компонента. Зависимость эвтектической температуры от толщины пленки висмута, находящейся между толстыми пленками германия, приведен на рис. 2б.

Исследования кристаллизации островковых пленок висмута на германиевой подложке проводились по методике, развитой в работах [10, 11] и основанной на изучении микроструктуры пленок, конденсированных на подложке с градиентом температур. В результате выполненных исследований получены значения предельной температуры $T_g = 451$ К, которая соответствует переходу от механизма конденсации пар → жидкость к механизму пар → кристалл, то есть температуре кристаллизации островков висмута на германиевой подложке. Соответствующая величина переохлаждения составляет $\Delta T = 93$ К ($\Delta T = T_s - T_g$).

Электронно-микроскопические снимки островковых конденсатов висмута на аморфных пленках германия при различных температурах приведены на рис. 3. Видно, что температура кристаллизации соответствует резкому изменению морфологической структуры пленок. Рис. 3в указывает на конденсацию по механизму пар → жидкость. На рис 3б, который соответствует граничной температуре $T_g$, конденсация проходила с изменением механизма, т.е. на начальной стадии роста островков висмута реализовался механизм конденсации пар → жидкость, а дальнейшее формирования пленки происходило уже по механизму пар → кристалл. На рис. 3а пленка имеет поликристаллическую структуру, являющуюся результатом конденсации по механизму пар → кристалл.

По электронно-микроскопическим снимкам профилей капель Bi на германиевой подложке при температуре 473 К методом наклонного наблюдения [12] определенна величина угла смачивания θ = 68° (рис. 4).

Как следует из многочисленных результатов исследований переохлаждения островковых пленок ряда металлов на различных подложках [10, 11] величина относительного переохлаждения $\Delta T/T_s$ зависит от краевого угла смачивания металлом подложки и достигает максимальных значений при θ ≥ 130°. В интервале углов 30° < θ < 130° зависимость $\Delta T/T_s = f(\theta)$ линейна для исследованных контактных систем. В соответствии с данными, приведенными в [10, 11], величина относительного переохлаждения при кристаллизации островков висмута на германиевой подложке при краевом угле смачивания 68° должна составлять 0,165, что хорошо согласуется с полученным в настоящей работе значением $\Delta T/T_s = 0{,}17$.

В соответствии с термодинамическим рассмотрением [8, 13] знак изменения температуры плавления частицы, заключенной в матрицу, определяется краевым углом смачивания ве-

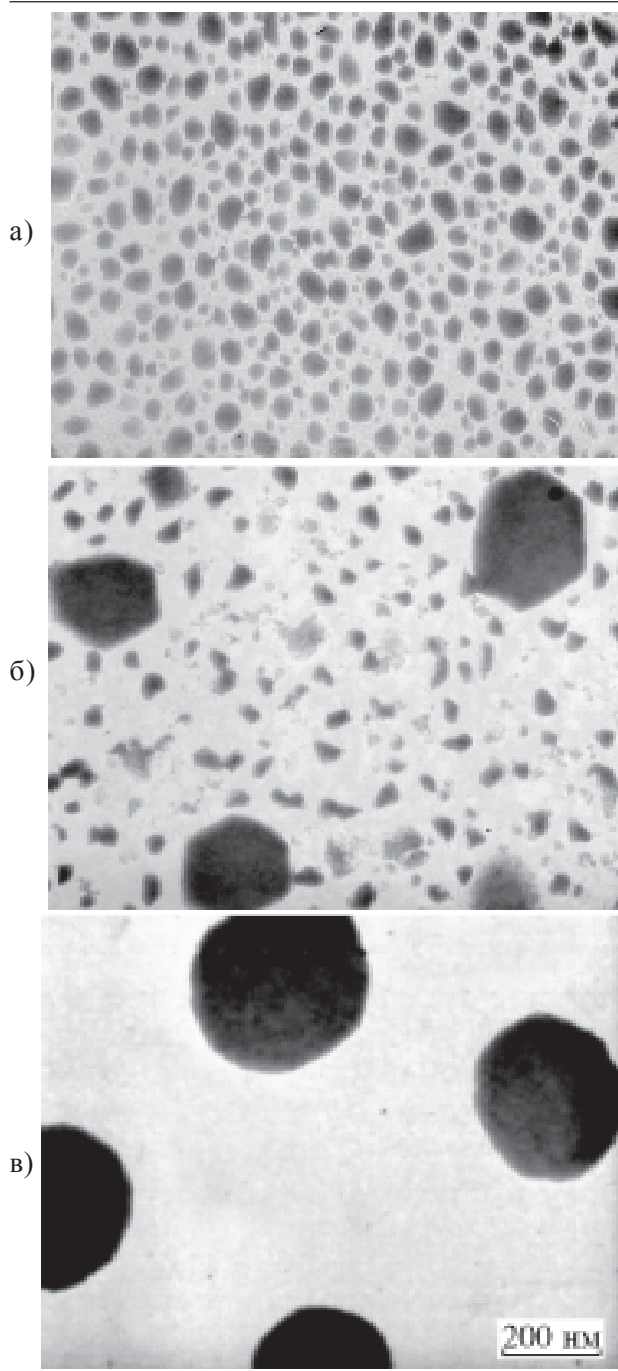

Рис. 3. Электронно-микроскопические снимки пленок Bi, конденсированных на германиевой подложке при различных температурах: а) $T = 434$ K, б) $T = 451$ K, в) $T = 473$ K.

ществом частицы материала матрицы. Для частицы в полностью несмачиваемой матрице, как и для свободной, должно наблюдаться понижение $T_s$ с уменьшением размера. В случае полного смачивания следует ожидать повышения температуры плавления с уменьшением характерного размера. Поскольку в исследуемой системе угол смачивания меньше 90°, то наблюдаемое понижение температуры плавления с уменьшением толщины пленки висмута между пленками гер-

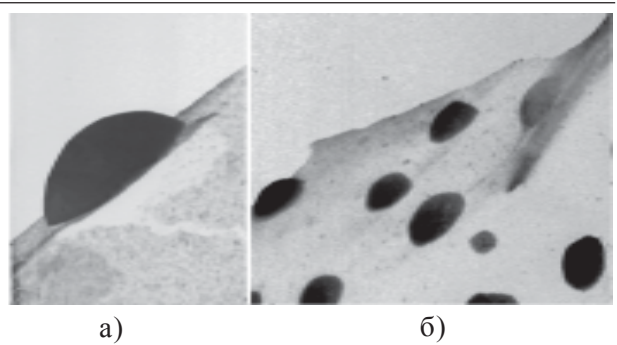

Рис. 4. Микроснимок частиц висмута на наклонных участках германиевой подложки (а): ×40000, б): ×20000).

мания значительно меньше, чем для свободных пленок или в случае полного несмачивания [14]. Однако, для объяснения наблюдаемых изменений температуры плавления, как показывает анализ полученных результатов и данных работ [2, 3, 8], недостаточно лишь рассмотрения процессов смачивания материалом частицы матрицы. Не­обходимо также учитывать тип соответствующей фазовой диаграммы и ее эволюцию при изменинии характерного размера. Для качественного объяснения наблюдаемого уменьшения температуры плавления от толщины тонких пленок между двумя толстыми пленками Ge в связи с соответствующими фазовыми диаграммами удобно воспользоваться методом геометрической термодинамики, основанным на учете роли поверхности и, следовательно, вклада свободной поверхностной энергии в общую свободную энергию [15]. Как следует из построения диаграммы эвтектического типа методом геометрической термодинамики, выполненного в работе [8], уменьшение характерного размера компонен­тов (или одного из них) приводит к смещению (в случае одного дисперсного компонента – к асимметричному смещению) линий фазовой диаграммы в область более низких температур по сравнению с диаграммой для массивных образцов.

Таким образом, полученные экспериментальные данные о снижении эвтектической температуры с уменьшением толщины пленок в совокупности с имеющимися в литературе результатами по снижению температуры плавления чистых компонентов и по увеличению растворимости, подтверждает эволюцию диаграммы состояния бинарных сплавов, которая проявляется в сдвиге ее в область более низких температур при переходе к высокодисперсным системам.

### ВЫВОДЫ

1. Исследование слоистых пленочных систем, создаваемых в вакууме путем последователь-

ной конденсации компонентов, показали, что температура плавления тонких пленок Bi, которые находятся между двумя толстыми пленками Ge, понижается с уменьшением их толщины.

2. Использование дифференциального метода регистрации температур плавления позволило с применением толстопленочных слоистых систем определить эвтектическую температуру для бинарной системы Bi аморфный Ge.

3. Установлено, что кристаллизация островковой пленки висмута на германиевой подложке происходит при температуре 451 К, что на $\Delta T = 93$ К ниже температуры плавления висмута в компактном состоянии.

**ПЛАВЛЕННЯ ТА КРИСТАЛІЗАЦІЯ В ШАРУВАТІЙ ПЛІВКОВІЙ СИСТЕМІ Ge-Bi**
**Богатиренко С.І., Гладких М.Т., Дукаров С.В., Криштал О.П.**

Приводяться результати досліджень плавлення і кристалізації в шаруватій плівковій системі вісмут – германій, яка створюється у вакуумі шляхом послідовної конденсації компонентів при випаровуванні їх з незалежних джерел. Показано, що температура плавлення в указаній системі знижується зі зменшенням товщини плівки легкоплавкого компонента. Диференціальний метод, використаний для реєстрації температури плавлення, дозволив знайти значення евтектичної температури $T_\varepsilon = 542$ K у системі Bi-Ge. Визначено величину переохолодження при кристалізації ($\Delta T = 93$ K) та крайовий кут змочування ($\theta = 68°$) для острівцевих плівок вісмуту на аморфній германієвій підкладці.

**MELTING AND CRYSTALLIZATION IN LAYERED FILM SYSTEM Ge-Bi**
**Bogatyrenko S.I., Gladkikh N.T., Dukarov S.V., Kryshtal O.P.**

The results of studies of melting and crystallization processes in Bi-Ge layered film system are presented. These systems were prepared by subsequent condensation of components in vacuum. It has been shown that the melting temperature in system under study decreases with the decrease of Bi film thickness. The differential technique used for melting temperature registration enables us to measure the value of eutectic temperature $T_\varepsilon = 542$ K in the system. The values of supercooling upon crystallization ($\Delta T = 93$ K) and wetting angle ($\theta = 68°$) have been determined for Bi islands on amorphous Ge substrate.